# Towards Cloud Efficiency with Large-scale Workload Characterization

*Anjaly Parayil[1], Jue Zhang[1], Xiaoting Qin[1], Íñigo Goiri[1], Lexiang Huang[1,2], Timothy Zhu[2], Chetan Bansal[1]*

[1]*Microsoft*     [2]*The Pennsylvania State University*

## Abstract

Cloud providers introduce features (*e.g.*, Spot VMs, Harvest VMs, and Burstable VMs) and optimizations (*e.g.*, oversubscription, auto-scaling, power harvesting, and overclocking) to improve efficiency and reliability. To effectively utilize these features, it's crucial to understand the characteristics of workloads running in the cloud. However, workload characteristics can be complex and depend on multiple signals, making manual characterization difficult and unscalable.

In this study, we conduct the first large-scale examination of first-party workloads at Microsoft to understand their characteristics. Through an empirical study, we aim to answer the following questions: (1) What are the critical workload characteristics that impact efficiency and reliability on cloud platforms? (2) How do these characteristics vary across different workloads? (3) How can cloud platforms leverage these insights to efficiently characterize all workloads at scale?

This study provides a deeper understanding of workload characteristics and their impact on cloud performance, which can aid in optimizing cloud services. Additionally, it identifies potential areas for future research.

## 1 Introduction

**Motivation.** Cloud platforms are a key component of the global IT infrastructure, with major players like Amazon Web Services [3], Microsoft Azure [41], and Google Cloud [27] providing a wide range of cloud services to individuals and organizations around the world. With the increase in cloud adoption, these providers are constantly offering new features and services, as well as introducing various optimizations to increase the efficiency and reliability of the platforms. These include new VM types (Spot VMs [4, 10, 18], Harvest VMs [7], Burstable VMs [58]), dedicated interfaces (*e.g.*, auto-scaling [40]), and optimizations that are internal to the cloud (*e.g.*, oversubscription [33], pre-provisioning [63]). However, to maximize the opportunities brought by these new features and cloud optimizations, it is necessary to understand the characteristics of workloads running in the cloud. Using this information, cloud platforms can determine which cloud workloads are most suitable for specific features or optimizations.

**Challenges.** Not all workload requirements are immediately evident from a single telemetry or workload signal. This limitation is prevalent in previous work, which primarily focuses on resource management [2, 23, 30, 68], workload profiling [14, 16, 20, 31, 37], and heuristic and Machine Learning (ML)-centric approaches [12]. These approaches rely on VM or container level telemetry to provide insights on resource utilization, deployment size, lifetime, and similar characteristics. VM or container-level telemetry may not provide complete information on workload requirements, such as delay sensitivity, which is crucial in determining whether a CPU needs to be overclocked [29]. While it is possible to ask the workload owner for this information, this approach is not scalable due to the large number of cloud workloads.

We also need to understand which workloads benefit from each cloud optimization. For example, overclocking the CPU running a memory-bound workload will not result in much performance improvement [29]. Even if we know that a cloud optimization impacts the performance of a workload, not all workload users care. For example, some users may care a lot about workloads finishing jobs as quickly as possible, but other users may not care about latency and would rather complete the jobs in the cheapest way possible, even if it takes longer.

**Our work.** We first identify the characteristics (*e.g.*, preemptibility or delay tolerance) required for various common cloud platform optimizations. Given that these characteristics are often not readily available, we conducted a survey of 188 first-party (*i.e.*, internal) workloads comprising over 100,000 VMs in Microsoft, a major public cloud provider. Through this empirical study, we categorize the workloads based on their functionality and find that the workloads can be grouped into six distinct classes. We study the performance, reliability, geographical, and scalability requirements of workloads

---

[1]Lexiang Huang was an intern at Microsoft.

across these classes.

Furthermore, we map each workload class to different cloud optimizations based on their suitability. We find that cloud providers can utilize the unique profiles of workload class and enable selected cloud optimizations to enhance their efficiency and reliability. For instance, the "Overclocking/Underclocking" optimization is very helpful for the performance of the "Web Proxy" class.

Our findings are also beneficial to cloud users, as they can help reduce overall costs while achieving desired performance. While our study focuses on first-party workloads of Microsoft, the methodology and insights generalize to other external workloads as well. This approach paves the way for an automatic characterization of cloud workloads at scale.

**Contributions.** In this paper, we make the following contributions:

- We present the first large-scale empirical study and characterization of real-world cloud workloads, with a focus on the relationship between workload characteristics and cloud optimizations.

- We derive insights on the workload characteristics and estimate opportunities for various cloud optimizations.

- We open-source the survey form used to collect information from the workload owners and aim to publish the raw data (pending the necessary privacy and security approvals).

The rest of the paper is organized as follows: Section 2 overviews some common cloud platform optimizations. Section 3 discusses the methodology of the survey. Section 4 analyzes workload characteristics and presents insights. Section 5 explores the implications derived from the study. Section 6 outlines potential future opportunities that arise from this work. Section 7 provides an overview of the related research. Section 8 presents our conclusions.

## 2 Cloud platform optimizations

We define a workload as a collection of applications, services, and data that support a specific process. To optimize cloud platforms, it's crucial to have a clear understanding of the unique requirements of each workload. Based on an extensive literature survey and discussions with domain experts working on cloud efficiency at Microsoft, we identify ten popular cloud optimizations and the necessary workload characteristics to apply them. Table 1 summarizes the workload characteristics required by each optimization.

*Auto-scaling.* To allow users to not always provision VMs for the peak load, cloud providers offer an auto-scaling feature to dynamically adjust the number of VMs based on load [40]. This requires workloads to be designed to support scale in and scale out. Providers offer auto-scaling as a separate service [6, 42] where users define an auto-scaling policy for their VMs. A workload that is either stateless or delay tolerant is suitable for auto-scaling if it has relaxed deployment time requirements.

*Spot VMs.* To monetize unallocated capacity, cloud providers offer VMs with relaxed SLOs at discounted prices. Specifically, they offer Spot VMs [4, 10, 18], which are low-priority VMs that would be evicted if their resources are needed by on-demand VMs. Providers offer Spot VMs through deployment flags and new VM types. Some providers offer dynamic pricing to decide which Spot VMs to evict first. Preemptible workloads are suitable for this optimization [5, 9].

*Harvest VMs.* Spot VMs are inefficient as they require creating and removing VMs to utilize all the resources in a server. Harvest VMs build on top of Spot VMs to dynamically grow and shrink a VM's size to utilize spare CPU [7, 59], memory [25], and storage [46] in the server. This is similar to Burstable VMs [8, 17], but gives providers more flexibility in determining when resources expand and shrink. Providers offer harvesting as a fixed VM type or a deployment flag specifying the amount of resources to harvest [7]. Harvest VMs are ideal for workloads that are both preemptible (scale in/out) and delay tolerant (scale up/down).

*Overclocking.* To improve the performance of workloads running in the cloud, providers can increase the CPU frequency of VMs [29] while considering the trade-offs with reliability and power budget. However, not all applications benefit from faster CPUs. A workload is suitable for overclocking if it is both delay-sensitive and non-preemptible, and it has high CPU utilization periods [29] (*i.e.*, $95^{th}$ percentile of maximum percentage CPU utilization greater than 40%). Cloud providers offer dedicated VM types that allow higher frequencies and provide interfaces for processor power states [49].

*Underclocking.* On the other hand, underclocking reduces the CPU frequency and decreases the power requirements at the cost of performance reduction. A workload is suitable for underclocking if it is either delay-tolerant, non-user-facing, or highly preemptible. Cloud providers offer dedicated VM types that allow for accounting of $CO_2$ and energy savings.

*VM pre-provisioning.* To reduce the time to create a VM, providers may provision VMs ahead of time and instantiate them as and when requested by workloads [64]. This is ideal for workloads with tight delay requirements for deploying new VMs, and it works well with auto-scaling as it adds VMs quickly when needed (*e.g.*, a load spike). However, when VMs are pre-provisioned, they are allocated a fixed amount of resources (such as CPU, memory, and storage) regardless of whether or not those resources are actually being used. Currently, cloud platforms infer which VMs to pre-provision without considering the utility to the workload. This can result in wasted resources and increased costs for the cloud provider and the user. Disabling VM pre-provisioning when not needed can save cost while have little impact on performance.

*Region-agnostic placement.* Some regions are cheaper and greener (*e.g.*, lower $CO_2$ emissions) than others. To reduce cost and emissions, workloads can run on VMs in different regions. This is ideal for workloads without latency or data-locality requirements that constrain them to a specific region. There are several proposals to do this semi-automatically [1, 52], but there are no commercial solutions yet, and workload owners currently need to manually specify the region.

*VM oversubscription.* Some VMs do not utilize all the resources assigned to them all the time. To increase server utilization, platforms can oversubscribe servers by placing more VMs [61]. This optimization relies on statistical multiplexing the utilization of different workloads. If all VMs spike in load at the same time, the platform throttles the least critical VMs. This is ideal for workloads with varying usage over time. A workload that is either delay-tolerant or non-user-facing is suitable for oversubscription if its $95^{th}$ percentile CPU utilization is less than 65% [20]. Cloud platforms usually infer which VMs can be oversubscribed and by how much [20].

*VM rightsizing.* Similarly to VM oversubscription, the cloud platform can identify underutilized VMs and recommend a different VM type that is smaller [11]. The cloud platform can proactively adjust the VM type based on workload characteristics. This is beneficial for workloads that can scale down and have lower utilization (*e.g.*, <50% utilization), allowing for a transition to a smaller VM size (typically half size).

*Multi-availability datacenters (MA DCs).* To reduce costs, cloud providers can remove redundant infrastructure for power delivery and cooling. During infrastructure failures or maintenance events, the cloud platform may need to throttle or even selectively turn off a subset of servers [65]. Multi-availability DCs take advantage of workloads that require low availability. Currently, the cloud platform infers which VMs are less critical and throttles them down or even evicts them if needed.

## 2.1 Relevant workload characteristics

There are many ways to characterize workloads, but only a subset relates to cloud optimizations. We identify the workload characteristics needed by cloud optimizations and categorize them into: (1) performance and reliability, (2) geographical, and (3) scalability.

**Performance.** Workloads have different performance requirements.

*User-facing.* Whether the workload is serving users and shows a periodic pattern.

*Delay tolerance.* Whether the workload tolerates delays (including deadlines). This can go from tail-latency SLOs ($< 200ms$) to batch job deadlines (finish a job before noon).

**Reliability.** Related to the workload availability requirements.

*Availability.* How much the workload allows a VM to not be available. This is usually measured in number of nines.

*Preemptibility.* How well the workload handles losing VMs. Most cloud workloads are built with fault tolerance in mind. However, there are different degrees and not all workloads support losing 50% of their VMs.

**Geographical.** Related to the workload requirement to run in a specific location or region.

*Proximity and locality.* Whether the workloads have location constraints. This can be related to latency (proximity to the users) or data locality (bandwidth requirements to access data in other locations). This indicates whether VMs running a workload can change locations at runtime.

**Scalability.** We define scalability as the ability of the workload to handle an increase in demand with consistent Quality-of-Service. A scalable workload can efficiently handle an increase in demand with more resources (scale in/out or scale up/down) [40]. The requirements to deploy additional resources for a workload are also crucial for better scaling.

*Scale up/down.* Whether the workload can adjust to changing the size of the VM. For example, switching to a VM with more or fewer cores or memory. This also applies to changes in CPU frequency.

*Scale out/in.* Whether the workload can adjust to the addition or removal of VMs.

*Deployment time.* The time that a workload requires to setup after getting a new VM. This relates to the scale out component.

## 3 Survey methodology

To understand the characteristics of workloads running in the cloud, we conduct a large-scale survey of first-party workloads at Microsoft. We design this survey based on the characteristics that existing cloud platform optimizations require to operate. This is the largest empirical study of the characteristics of workloads running in the cloud to date.

**Scope.** At Microsoft, first-party workloads consist of internal services for research and development, infrastructure management, and first-party services like communication, gaming, and data management that are offered to third-party customers. The company tracks its first-party workloads in an internal directory, which is categorized into different divisions based on the organizational hierarchy. In Table **??**, we summarize all 14 divisions at Microsoft along with relevant metrics. We select all the workloads with non-zero core usage from the week of June $20^{th}$-$27^{th}$ 2022 to calculate the relevant metrics. Division 1 and 2 are the largest divisions in terms of the number of VMs, cores, and deployed regions. Division 1 mainly encompasses internal workloads for research and development, as well as tools and platforms for cloud services offered by Microsoft. Division 2 is responsible for creating innovative products and services for people and organizations. This division encompasses a wide range of workloads, including web search, collaboration and productivity suites, and real-time communication services. In addition to this, we also have easier access to the workload owners from this division while having representative features. Therefore, we target Division 2 for this study.

| Cloud optimization | Required workload characteristics |
| --- | --- |
| Auto-scaling | (Stateless ∨ Delay tolerant) ∧ Relaxed deployment time |
| Spot VMs | Preemptible |
| Harvest VMs | Preemptible ∧ Delay tolerant |
| Overclock | Delay sensitive ∧ Non preemptible ∧ P95 Max % CPU > 40 |
| Underclock | Delay tolerant ∨ Non user-facing ∨ Preemptible |
| Pre-provision | Stringent deployment time |
| Region-agnostic | Geographical requirements |
| VM oversubscription | (Delay tolerant ∨ Non user-facing) ∧ (P95 Max % CPU < 65) |
| VM rightsizing | P95 Avg % CPU < 25 ∧ P95 Avg % mem < 25 ∧ P95 Max % CPU < 50 ∧ P95 Max % mem < 50 |
| Multi-availability DC | Low availability |

Table 1: Popular cloud platform optimizations and their relevant workload characteristics (∨: or, ∧: and, P$x$ Max: $x^{th}$ percentile of maximum utilization, P$x$ Avg: $x^{th}$ percentile of average utilization).

Overall, we consider 1034 workloads among which the largest one employs over 100k VMs. These workloads are deployed to 49 regions across the world and used by hundreds of millions of users.

**Process.** We initially conducted three in-person interviews to refine, clarify, and disambiguate survey questions, whenever necessary. To further evaluate the response rate and ambiguity of questions, we shared an online form with another 27 workload owners selected via weighted random sampling (weighted by their core usage). After refining the survey based on the answers for these first 30 workloads, we sent the online form to the remaining 1004 services. The final response rate of the survey is 19%, which covers workloads (24.3% of the total cores).

**Questions.** Based on the requirements of the optimizations described in Section 2, we formulate twenty-one questions to collect information on workload characteristics and requirements. We start asking for a high-level description including: (1) the main functionality, (2) links to documentation, (3) if it is composed by other workloads, and (4) if the workload is user-facing or not. Then, we move into specific questions in the three categories identified in Section 2.1. Here, we give a summary of the questions in each category, and interested readers may refer to Appendix A for the complete list of survey questions.

*Performance.* We ask if the workload is tolerant to delays and if they have some performance expectation (*e.g.*, latency). We use the expected latency values provided by the workload owners to verify their response to delay tolerance.

*Reliability.* We ask for availability and fault tolerance requirements.

*Geographical.* We ask whether the workload can be migrated to other regions. We also ask for the factors restricting cross geo-migrations if their workloads cannot be migrated to other regions.

*Scalability.* We ask for the potential to run in a different VM configuration (*e.g.*, running in more/less VMs). We also ask if the workload is stateless as stateless workloads are potential candidates for easy scaling due to their input-independent control flows [66].

**Free-form text responses.** We also include questions with free-form text responses to get more insights. We use the open coding approach [54] to categorize these free-form responses. We first randomize the data and split it into 3 sets: (1) taxonomy set (30% of the total data): to initially assign classes based on what annotators perceive as the most appropriate labels, (2) validation set (20%): to make sure no new labels emerge, and (3) label set (remaining 50%): to evaluate if the annotators agree on the labels, and we use an inter-annotator score [19] to identify that.

We assign free-form text responses to two annotators who label the taxonomy set independently. Annotators are authors of this work with extensive background in systems design and modeling. Subsequently, they discuss the categories and reach a consensus. Next, they independently label the validation set to make sure no new categories emerge. Then, they have another discussion to settle disagreements and define a common understanding of each category. Lastly, they annotate the label set and compute the inter-annotator agreement score using Cohen's kappa [19]. Annotators use the resulting score to identify any disagreements. With this approach, the annotators settle all disagreements and create a systematically labeled dataset for further analysis.

**Generality.** To evaluate the generality of the results, we compare the general characteristics of the sample workloads in our survey with the rest of the division and across divisions. Table 2 shows the corresponding values. This data indicate that the characteristics of the workloads in the sample are very similar to the rest of the division and the company and that our survey samples are representative. Moreover, Table ?? supports our conclusion that the selected division is a good representation of Microsoft.

## 4 Survey results

In this section, we present the results from the survey. We start by describing the workloads at a high level and move into the detailed results for each of the workload characteristics categories. Unless stated otherwise, when talking about percentages, we are referring to CPU core percentages.

### 4.1 Workload classes

To provide high-level intuitions about the results, we categorize the workloads into major classes. We correlate the descriptions provided by the workload owners in the survey with their internal directory information and use open-coding to group these descriptions into six classes. The annotators have near-perfect agreement (*i.e.*, Cohen's kappa score of 0.965). The disagreements were mostly due to workloads fitting into multiple classes and resolved them by adopting the most generic one and providing clearer descriptions.

Based on these classes, we study their popularity with respect to their frequency and core usage. Table 3 shows these classes, their description, core usage, and frequency. "Web Apps" and "Big Data" constitute 50% and 18% of all workloads, respectively. On the other hand, "Big Data", "Web Apps", and "RTC" workloads consume most of the cores (83%). Despite having lower frequency, "RTC" and "ML Inference" workloads exhibit significantly higher average core usage, which aligns with their compute-intensive nature. Note that we specifically include an "ML Inference" category in the taxonomy, as opposed to "ML Training" which is performed offline in dedicated GPU clusters (outside the scope of this study).

To validate the generality of the survey, we also manually labeled the remaining workloads from the division. Table 4 displays the distribution of workload classes in both the survey and the entire division. The distributions exhibit similar patterns, with "Web Apps" being the most common class, followed by "Big Data" and "DevOps".

> 📝 **Takeaway**
> Diverse workloads can be categorized into six main classes, out of which "Big Data", "Web Apps", and "RTC" accounts for the majority of core usage and thus

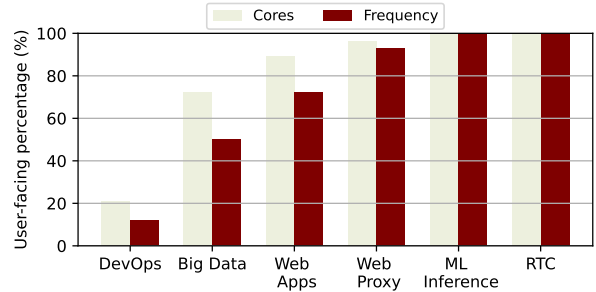

Figure 1: User-facing nature per workload class.

> should be the primary targets for implementing cloud optimizations.

### 4.2 Performance requirements

**User-facing.** These are workloads which serve real-time traffic which expects a live human interaction (*e.g.*, web applications composed of front-end web servers and databases). These workloads typically exhibit periodic utilization patterns, with high activity during the day and low activity at night. They are commonly referred to as performance-critical workloads [33].

Figure 1 shows the distribution of user-facing nature with respect to workload class in terms of their frequency and percentage core usage. The "RTC" and "ML Inference" classes are completely user-facing. This is intuitive as "RTC" handles mostly online meetings and the majority of the "ML Inference" workloads support interactive functions (*e.g.*, text prediction, smart reply). As expected, the percentage of user-facing workloads is also high among "Web Apps" and "Web Proxy" classes.

"DevOps" contains the lowest number of user-facing workloads, which is also highly intuitive as they are mainly used for software development and IT operations. "Big Data" contains a similar portion of user-facing and non-user-facing workloads. This is due to the presence of diverse workloads such as data analytics, hot data-stores, and event hub workloads.

> 📝 **Takeaway**
> Considering the user-facing nature, "RTC" and "ML Inference" have high performance requirements while "DevOps" workloads are usually less critical.

**Delay tolerance.** Delay tolerant workloads have the flexibility to tolerate delays within a specified deadline, allowing for better resource management at the cost of increased service delays [60]. Overall, around 56.4% of the workloads are delay sensitive and the remaining 43.6% are delay tolerant.

|  | Survey | Selected division | Overall |
|---|---|---|---|
| Number of workloads | 188 | 1034 | 4275 |
| Number of regions | 49 | 49 | 55 |
| P50 & P99 Avg CPU | [5.5%, 91.0%] | [5.2%, 95.7%] | [5.6%, 74.9%] |
| P50 & P99 Max CPU | [35.2%, 99.2%] | [28.2%, 99.9%] | [18.7%, 99.2%] |
| P50 & P99 Avg Memory | [30.1%, 100.0%] | [29.9%, 100.0%] | [31.2%, 100.0%] |
| P50 & P99 Max Memory | [39.1%, 100.0%] | [37.9%, 100.0%] | [40.4%, 100.0%] |
| P50 & P99 #VMs | [353, 34K] | [43, 344K] | [12K, 29M] |
| P50 & P99 VMs lifetime (h) | [646, 23,320.42] | [4, 18,751.72] | [0, 21,411] |

Table 2: Comparison of characteristics of the workloads in the survey, selection division, and overall at Microsoft.

| Class | Description | Cores | Frequency |
|---|---|---|---|
| Big Data | Database management operations (*e.g.*, ingestion, update) and various analytical (*e.g.*, descriptive, diagnostic, predictive, prescriptive, and cognitive) tasks over large or complex databases. It includes pipelines and event hubs that receive and process a large number of events per second. | 32.4% | 18.1% |
| Web Apps | Applications with a front-end, services with external APIs, and backend infrastructure that supports other applications. For example, a product offered by the cloud provider using an online interface. | 27.3% | 50.0% |
| Real-Time Comm (RTC) | Audio and video workloads (*e.g.*, online meetings) that support real-time communication. | 24.1% | 6.4% |
| ML Inference | Inference of machine learning models. | 11.0% | 4.8% |
| Web Proxy | Intermediary between a client application and the server. Examples are relay, proxy, and gateway. | 3.9% | 7.4% |
| DevOps | Tools (not part of final offering) for software development and IT operations. It covers code development, phased deployment (*e.g.*, staging, pre-production), and pipelines for continuous integration. Examples are internal workloads for testing and tools to accelerate engineers' debugging. | 1.3% | 13.3% |

Table 3: Workload class descriptions with their core usage and frequencies.

| Workload Class | Survey | Division |
|---|---|---|
| Web Apps | 50.0% | 50.9% |
| Big Data | 18.0% | 19.1% |
| DevOps | 13.3% | 17.0% |
| Web Proxy | 7.4% | 6.0% |
| RTC | 6.4% | 4.2% |
| ML Inference | 4.8% | 2.7% |

Table 4: Generality for the workload classes.

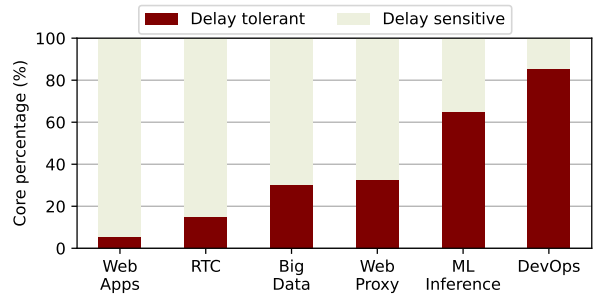

Figure 2: Delay tolerance requirements per workload class.

However, it is worth noting that delay-sensitive workloads account for the majority of the cores (75.5%). Practically, workloads do not usually leverage delay-sensitivity information, resulting in users spending more money and cloud platforms dedicating more resources than necessary.

Figure 2 shows the delay tolerance across workload classes. More than 80% of the cores from the "DevOps" class are delay tolerant. A substantial number of cores from "RTC" and "Web Apps" are delay sensitive. This is expected as they are mostly user-facing workloads. For these workloads, the expected latency values are typically in the range of milliseconds to a few seconds. On the other extreme, workloads like "Big Data" require more time to process and analyze large amounts of data. For these workloads, the targeted latencies are typically in the range of seconds to minutes. There is a large fraction of delay-tolerant cores (over 60%) in the "ML Inference" class because some workloads provide user-delight features (e.g., text prediction) for which users have a larger tolerance for service performance.

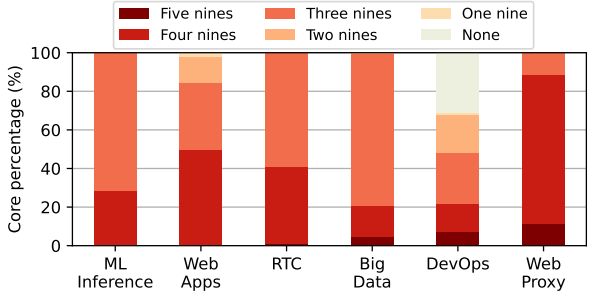

(a) Availability.

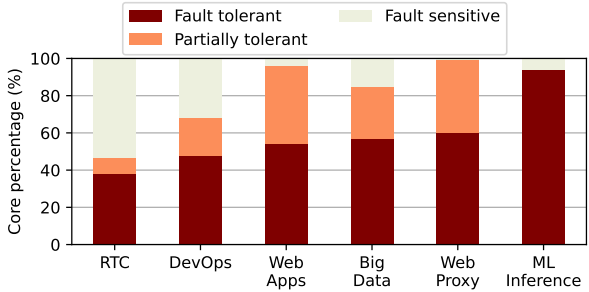

(b) Fault tolerance.

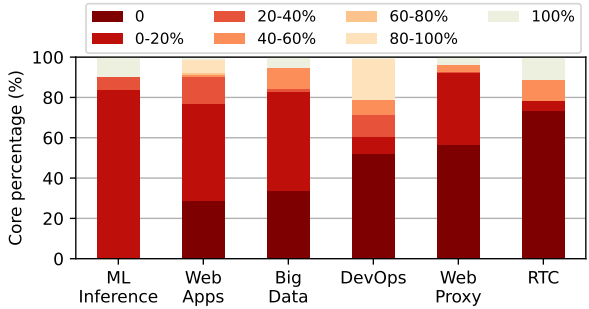

(c) Preemptibility.

Figure 3: Reliability requirements per workload class.

> **Takeaway**
> There is a significant opportunity to optimize the execution of predominantly delay-tolerant workloads (*e.g.*, "DevOps") for both workload owners and the cloud platform, such as through the use of Harvest VMs.

## 4.3 Reliability requirements

**Availability.** We first study the availability requirements [51, 57]. A highly available workload remains operational with minimal downtime in the event of a disruption (e.g., hardware failure, networking problems). An availability requirement of "five nines", "four nines", "three nines", "two nines", and "one nine" represent a downtime of 5 minutes, 52 minutes, 8.77

| #Nines | Cores | Frequency |
|---|---|---|
| Five | 2.4% | 10.3% |
| Four | 34.5% | 25.3% |
| Three | 58.0% | 35.6% |
| Two | 4.0% | 14.4% |
| One | 0.5% | 6.2% |
| None | 0.4% | 8.2% |

Table 5: Availability requirements.

hours, 3.65 days, and 36.53 days per year, respectively [13].

Table 5 shows that most workloads require an availability of "three" or "four nines". Only 10% of these workloads expect an availability of "five nines", with their daily core usage accounting for only 2.4% of the total daily usage. Availability is an important consideration for most workloads, given that only 8.2% of the workloads have no availability requirements.

Figure 3a shows the availability requirements for each workload class. The cores from the "Web Proxy", "ML Inference", and "RTC" classes expect an availability of at least "three nines". This could be attributed to the higher percentage of user-facing workloads in these classes (cf. Figure 1). This is intuitive as "Web Proxy" workloads are often critical as they serve intermediate components between the client and server. The requirements for "DevOps" range from "one nine" to "five nines", and there is a significant percentage without any availability requirements (31.4%). This class usually contains workloads for software development and IT operations that are less critical to the functioning of the overall system. "DevOps" workloads may be more amenable to optimization and resource sharing than other types of workloads, since there is more flexibility in terms of when and how these tasks are run.

> **Takeaway**
> While a significant percentage of cores (95%) requires an availability greater or equal than "three nines", only a limited percentage require "five nines". "DevOps" workloads offer more flexibility in terms of when and how to run those workloads.

**Fault tolerance.** A fault tolerant workload can handle VM failures without impacting the service [34, 57]. A workload is fault-tolerant if its performance is insensitive to system failures (*e.g.*, a VM failure does not impact the operation or latency of a workload) [34]. Partially fault-tolerant workloads contain certain components that are tolerant to faults.

Table 6 shows the distribution of fault tolerance among workloads. More than 50% of the workloads are completely fault tolerant. Specifically, "ML Inference" seems to be mostly fault tolerant (93.7%). This is because most of these workloads use frameworks that handle fault tolerance and re-route

| Workload characteristic | Cores | Frequency |
|---|---|---|
| Fault tolerant | 55.7% | 53.7% |
| Partially fault tolerant | 24.2% | 23.9% |
| Fault sensitive | 20.2% | 22.3% |

Table 6: Fault tolerance requirements.

| | Cores | Frequency |
|---|---|---|
| 0% | 39.3% | 26.3% |
| 0-20% | 41.1% | 22.2% |
| 20-40% | 4.8% | 12.4% |
| 40-60% | 6.5% | 16.0% |
| 60-80% | 0.3% | 5.2% |
| 80-100% | 1.8% | 4.6% |
| 100% | 6.1% | 13.4% |

Table 7: Preemptibility requirements.

| Workload Characteristic | Cores | Frequency |
|---|---|---|
| Region-agnostic | 47.5% | 58.0% |
| Partially region-agnostic | 13.9% | 19.1% |
| Not region-agnostic | 38.6% | 22.9% |

Table 8: Geographical requirements.

queries. Workloads that are partially fault tolerant contribute to 24.2% of the total core usage.

Figure 3b shows the breakdown of fault tolerant cores with respect to workloads classes. Workloads that are sensitive to faults account for only 20.2% of the total core usage. "RTC" (53.7%) and "DevOps" workloads (31.8%) use a significant percentage of fault-sensitive cores. "RTC" workloads are part of communication system, which is latency sensitive, and this could be a potential reason for the presence of a significant percentage of fault sensitive cores.

> **Takeaway**
> Cores from the "ML Inference" class are highly fault tolerant, and this workload can remain operational with minimal downtime or data loss in the event of a disruption, whereas the "RTC" class contains the most percentage of fault-sensitive cores.

**Preemptibility.** While availability and fault tolerance relate to failures, preemptions are triggered by the cloud platform for efficiency reasons, so we consider preemptibility separately [21]. This refers to the ability to handle the loss of VM instances under normal operating conditions. For example, some workloads require all VMs to be up all the time (*i.e.*, 0% preemptibility) while others may allow 60% of the VMs to be running.

Table 7 shows the distribution of preemptible workloads. Around 60% of the workloads are preemptible by at least 0-20%. In addition, there is a small percentage of workloads that are 100% preemptible (13.4%). Highly preemptible services use significantly fewer cores than services with lower preemptibility. This suggests that these services are less complex or have fewer data dependencies.

Figure 3c shows the preemptibility requirements with respect to the workloads classes. "RTC" contains the highest number of cores (73.6%) with 0% preemptibility requirements. 84.1% of the "ML Inference" cores are at least 0-20% preemptible. On the other hand, "DevOps" shows diverse levels of preemptibility requirements that span from 0% to 80-100%. The higher percentage of "DevOps" cores with no preemptibility is due to their need to persist data as a part of streamlined packages and environments. Similarly, for "Web Proxy", most of the cores show preemptibility levels of 0% and 0-20%. This might be because these workloads serve as intermediate components between client and server and are often critical. "Web Apps" have a heterogeneous mix as it has a relatively high percentage of services with a preemptibility level of 0-20%, but also a relatively high percentage of services with a preemptibility level of 80% to 100%. This suggests that the preemtibility of "Web Apps" depends on the specific workload.

> **Takeaway**
> A good amount of "DevOps" and "ML Inference" workloads are suitable candidates to deploy on preemptible cores, while "Web Proxy" and "RTC" workloads are the last choice for preemptible cores.

### 4.4 Geographical requirements

We analyze the geographical requirements of various workloads and the factors restricting cross geo-migrations (*i.e.*, migrations spanning geographical regions). Region-agnostic workloads (RAW) can be deployed or migrated at least to any other region within a certain geo-locale without any negative impact on its operation. A workload is partially RAW if a certain percentage can be migrated or contains migratable components. These workloads are potential candidates for geographical load balancing during capacity shortages [1, 52].

Table 8 shows that fully RAWs (*i.e.*, without any proximity and location constraints) represent over 50% of the workloads. Fully and partially RAWs contribute to more than 60% of the cores.

**Factors limiting cross-region migration.** To categorize these factors, we use open coding and categorize the free-text responses from the workload owners. Cohen's kappa score of

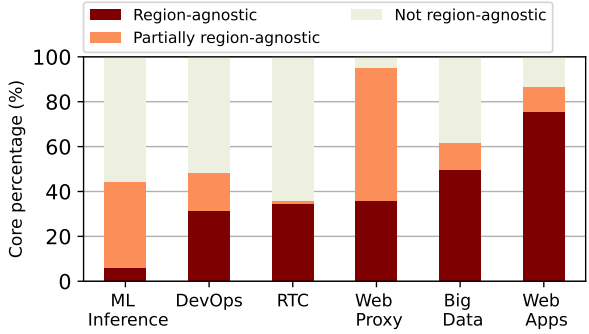

(a) Core usage of region-agnostic workloads.

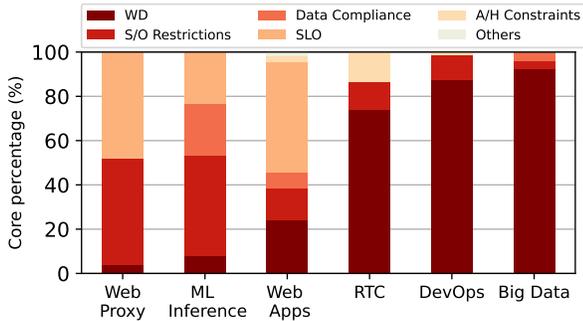

(b) Factors affecting cross geo-migrations.

Figure 4: Location constraints per workload class.

0.93 shows near-perfect agreement among the annotators. Disagreements were mostly due to insufficient clarity in the responses of workload owners. Annotators resolved those disagreements by reviewing all the available documentation on the workloads and adopting the appropriate labels.

Table 9 summarizes the five classes obtained from open-coding and the associated definitions of each class. Some of the workloads are restricted by a single factor whereas others are restricted by a combination of them. The most frequent combinations are ("Workload Dependencies", "Security/Offering restrictions") and ("Workload Dependencies", "Architecture/Hardware"). To better understand the influence of each category, we consider each category independently, perform normalization, and calculate their core distribution and frequency. "Workload Dependencies" (51.9%) and "Security/Offering Restrictions" (21.3%) are the most common factors restricting cross geo-migrations. "Others" includes unclear responses and accounts for only 0.2% of the cores.

Figure 4 shows the region dependency for each workload class as well as the factors restricting them from cross geo-migrations. Some observations are as follows:

**Web Apps** dominate the region-agnostic cores (75.4%). This is due to their inherent nature of responding to ad-hoc requests from end users, without the need to store request state. They also rely on components, such as web servers, application servers, and databases, that can be distributed across multiple geographic locations to improve performance and availability. Due to the user-facing nature of these workloads, the major factor restriction is due to "SLO" (*e.g.*, round trip time and latency).

**Big Data** contains nearly an equal amount of region-agnostic, partial, and region-affine cores. Certain components of "Big Data" workloads such as data storage or processing components are more sensitive to geographic location than others and require low latency or high bandwidth connections to other components. The cross geo-migration here mainly depends on "Workload Dependency" as they often have data dependencies.

**Web Proxy** comprises mostly fully or partially region-agnostic cores. This is because although proxies are user-facing with "SLO" restrictions and data privacy concerns, a portion of its cores can be migrated due to their stateless and lack of dependency on local resources (c.f. Figure 5a).

**RTC** contains the highest percentage of region-affine cores and the major restricting factors are "Workload Dependencies" and "A/H constraints". These workloads are generally composed of multiple microservices and are compute heavy.

**DevOps** is restricted because of "Workload Dependencies" and "S/O restrictions" (similar to "RTC"). These restrictions are mainly due to the packages and environments often needed for "DevOps" workloads.

**ML Inference** contains the lowest percentage of region-agnostic cores. The major restricting factors are related to data privacy and security ("S/O" and "Data Compliance"). They also often offer interactive inference with location specific features and the cost of spinning them up in a new region is often too high.

> **Takeaway**
> A major portion of cores (61.4%) could be completely or partially deployed to other geo-locations. "Web Apps" workloads are potential candidates for cross geo-migrations. Overall, besides "SLO", "Workload dependency" and "Security/Offering" are more dominant restricting factors for cross geo-migrations.

## 4.5 Scalability requirements

**Stateless vs stateful.** Stateless workloads can be easily scaled in/out as they require no state to be persisted. "Partially stateless" workloads contain both stateless and stateful components. For example, a partially stateless "Big Data" workload from the survey uses stateless components for data querying and stateful ones for collecting data and uploading it to a cloud storage. Table 10 shows that over 50% of the workloads are stateless and their core usage amounts to 45.5%.

| Categories | Description | Cores | Frequency |
|---|---|---|---|
| Workload Dependencies (WD) | Upstream and downstream dependencies associated with a workload, local cache requirements or storage affinities. | 51.9% | 48.2% |
| Security/Offering (S/O) | Related to security or due to differences in features offered in different regions. Examples include restrictions on IP ranges. | 21.3% | 25.9% |
| Service Level Objectives (SLO) | Restrictions such as latency, RTT requirements, and performance requirements. | 14.5% | 8.2% |
| Data Compliance (DC) | Region-specific data handling standards. Examples include EU DB restrictions. | 7.3% | 10.6% |
| Architecture/Hardware (A/H) | Imposed by underlying fabric and deployment framework. | 4.7% | 4.7% |
| Others | Factors not belonging to any of the defined classes and unclear responses. | 0.2% | 2.35% |

Table 9: Factors limiting cross geo-region migrations and their descriptions.

| Workload Characteristic | Cores | Frequency |
|---|---|---|
| Stateless | 45.5% | 51.6% |
| Partially stateless | 17.4% | 16.0% |
| Stateful | 37.1% | 32.4% |

Table 10: State requirements.

Figure 5a shows the breakdown of statefulness for each workload class. "Web Apps" and "Web Proxy" constitute most of the stateless cores. Most "Web Apps" workloads are stateless by nature as requests are "independent" (*e.g.*, responding to ad-hoc requests from end users). The high fraction of stateless cores in the "Web Proxy" class may be attributed to its main role of relaying messages, which often does not require storing any state.

The presence of stateless cores among "DevOps" and "RTC" workloads is significantly lower. "DevOps" workloads typically rely on a complex build environment with multiple steps and is thus more likely to be stateful. "RTC" is based on streaming data, which requires keeping significant state. These services are often designed to be more tightly coupled and stateful to support real-time communication or other complex workflows.

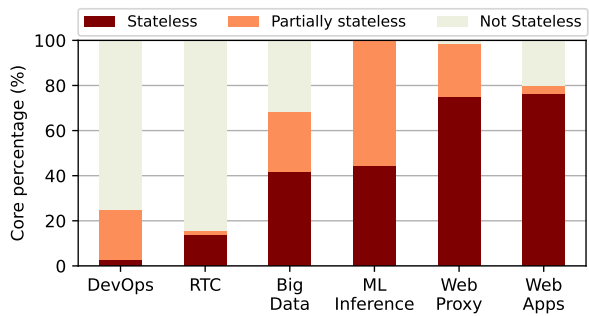

(a) Stateless vs. stateful.

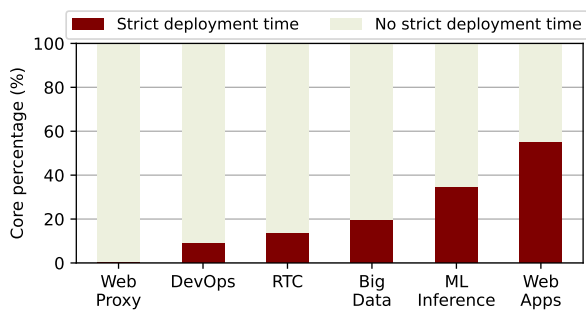

(b) Deployment time requirements.

Figure 5: Scalability requirements per workload class.

> 📝 **Takeaway**
> Stateless cores are prevalent in "Web Apps" and "Web Proxy" workloads, making them more flexible for scaling in/out. On the contrary, "DevOps" and "RTC" workloads are mostly stateful.

**Deployment time.** We define the deployment time requirements of a workload as strict if the VMs need to be deployed in less than one minute. Requirements greater than a minute are considered relaxed. Table 11 reports the distribution of deployment time requirements among workloads. Most workloads (more than 70%) have relaxed deployment time requirements.

| Deployment time | Cores | Frequency |
|---|---|---|
| Strict | 28.5% | 20.7% |
| Relaxed | 71.5% | 79.3% |

Table 11: Deployment time requirements.

Figure 5b shows the breakdown of deployment time requirements by workload class. Most workload classes, except for "Web Apps", have a low percentage of cores with strict deployment time requirements. The "Web Proxy" class contains the lowest percentage of cores with strict deployment time requirements (0.3%). "Web Proxy" workloads exhibit a better ability to handle an increase in demand with consistent QoS due to their stateless nature and relaxed deployment time requirements.

> **Takeaway**
> While most of the workloads don't have strict deployment time requirements, the majority of cores of the "Web Apps" class have a strict deployment time requirement, which should be carefully met when scaling out on demand.

## 5 Implications

### 5.1 Opportunities for cloud optimizations

**Relaxed requirements.** In this section, we use Table 1 to identify suitable platform optimizations for different workloads. We focus on workloads with relaxed requirements and enable specific cloud platform optimizations to enhance their performance. For example, we consider availability requirements lower than "four nines" to be low and preemptibility larger than 20% to be high. Table 12 shows the percentage of cores with relaxed characteristics for each workload class. This table highlights that each class has a unique profile. For instance, "Web Apps" are stateless, sensitive to delays, have low preemptibility, are flexible for cross geo-migrations, and have strict deployment time requirements.

**Results.** We use the cloud optimizations listed in Table 1 to calculate the percentage of cores from each workload class that are suitable for different cloud optimizations. For example, Table 1 suggests that workloads with high preemptibility and low availability requirements are suitable for "Harvest VMs". Therefore, we identify the cores with high preemptibility and low availability for "Harvest VMs" in each workload class to determine the percentage of suitable cores. Table 13 summarizes the results. From Table 13, we can see that there are significant opportunities for "Auto-scaling", "Spot VMs", "Underclocking", "Overclocking", "Region-agnostic" deployments, and "MA datacenters". Cloud providers can leverage these optimizations on applicable workload classes to monetize their unallocated capacity, reduce costs, and lower $CO_2$ emissions. Meanwhile, cloud users can consider the common characteristics within their workload class to identify optimizations that may lower their cost and/or improve performance and reliability. The limited opportunities for "Harvest VMs" and "VM Oversubscription" is due to the dominance of delay sensitive cores. Making workloads less delay sensitive can expand opportunities for these optimizations. It's worth noting that some workload classes show lower opportunities than others for certain cloud optimizations. We now discuss the particularities for each class:

**Web Apps** Most of these workloads can be deployed across multiple regions and take advantage of lower cost resources, reduce $CO_2$ emissions, and the impact of regional outages. They can be throttled during infrastructure failures or maintenance events (*i.e.*, "MA DC"). A significant percentage can be deployed on cost-effective "Spot VMs" for lower costs. "Auto-scaling" is another feature that can be used to optimize the its efficiency. "Overclocking" and "Underclocking" can also be used to improve workload performance and reliability.

**Big Data** Most of these workloads do not require "Pre-provisioning" ($\sim$ 74%) and over 70% of cores can enable "MA DCs". These workloads can also leverage "Underclocking/Overclocking". Similarly, 30% of the cores are suitable for "Region-agnostic" deployments.

**DevOps** A significant portion of these workloads can benefit from "Auto-scaling", "MA DCs", and do not require the "Pre-provisioning" of resources. Additionally, "Underclocking" can help reduce power consumption and save costs, which is especially important for "DevOps" teams that manage large-scale infrastructure.

**Web Proxy** Cloud providers can enable "Auto-scaling" for most of these workloads to reduce over-provisioning of VMs without the need for "Pre-provisioning". "Overclocking" and "Underclocking" can also be used to improve workload performance and reliability. These workloads are also a good target for "Rightsizing".

**RTC** "Overclocking" can enhance the performance and reliability of these workloads. Even small improvements in performance can have a significant impact on user experience. These workloads do not require pre-provisioning and can benefit from the use of "MA DCs" to improve reliability and reduce the impact of regional outages or performance issues.

| Relaxed requirements | Web Apps | Big Data | DevOps | Web Proxy | RTC | ML Inference | Overall |
|---|---|---|---|---|---|---|---|
| User-facing | 89.1% | 72.3% | 20.8% | 96.4% | 100% | 100% | 86.9% |
| No proximity & location constraints | 75.3% | 49.6% | 31.6% | 36.1% | 34.7% | 6.2% | 47.5% |
| Delay tolerant | 5.7% | 30.2% | 85.6% | 32.5% | 15.1% | 65.2% | 24.5% |
| Low availability (< Four nines) | 50.4% | 79.2% | 78.1% | 11.1% | 59.3% | 71.3% | 63.0% |
| High preemptibility (> 20%) | 23.1% | 17.0% | 39.5% | 7.6% | 21.6% | 15.9% | 19.6% |
| Fault tolerant | 54.3% | 57.0% | 47.4% | 60.0% | 37.8% | 93.8% | 55.7% |
| Stateless | 76.4% | 43.1% | 2.9% | 74.8% | 3.7% | 44.5% | 45.5% |
| No deployment time requirements | 45.1% | 74.4% | 98.5% | 99.6% | 86.2% | 63.0% | 68.8% |

Table 12: Percentage of cores with relaxed workload characteristics for each class.

| | Web Apps | Big Data | DevOps | Web Proxy | RTC | ML Inference | Overall |
|---|---|---|---|---|---|---|---|
| Auto-scaling | **26.2%** | **34.4%** | 88.1% | 95.6% | 13.0% | 63.0% | 33.1% |
| Spot VMs | **23.1%** | **22.0%** | 39.6% | 7.6% | **22.0%** | 19.0% | **21.6%** |
| Harvest VMs | 3.4% | 2.8% | 27.9% | 0.6% | 10.5% | 12.0% | 6.4% |
| Overclocking | **26.6%** | **45.4%** | 0.3% | 56.0% | 72.0% | 0.0% | **41.0%** |
| Underclocking | **34.4%** | **27.9%** | 99.0% | 39.8% | **26.9%** | 70.0% | **36.0%** |
| Non pre-provision | **45.1%** | 74.4% | 98.5% | 99.6% | 86.2% | 63.0% | 68.8% |
| Region-agnostic | 75.3% | **31.7%** | **31.1%** | 36.4% | **35.2%** | 6.94% | **43.0%** |
| VM oversubscription | 3.3% | 1.6% | 19.3% | **24.2%** | 13.1% | 12.2% | 7.6% |
| MA datacenters | 51.0% | 72.0% | 85.1% | 11.2% | 59.0% | 70.0% | 59.6% |
| VM rightsizing | 0.74% | 0.0% | 0.0% | **23.1%** | 0.0% | 7.3% | 2.1% |

Table 13: Percentage of cores suitable for each cloud platform optimization. Bold and underlined values indicate percentage cores greater than 20 and 50, respectively.

**ML Inference** A substantial portion of the cores from these workloads are suitable for "Auto-scaling" and do not need "Pre-provision". These workloads also enable early throttling in the event of failures with "MA DCs" and can benefit from "Underclocking".

**Takeaway**
Since different workload classes benefit from different sets of cloud optimizations, identifying the workload class is a good start at determining what optimizations to enable. This helps achieve greater cloud efficiency and reliability while minimizing users' costs.

## 5.2 Correlation between characteristics

Our empirical study in Section 4 shows that different workload classes often have different characteristics, indicating that workload class can be an important feature for determining workload characteristics. Cloud platforms can utilize these characteristics to further improve the efficiency and reliability. To analyze how different characteristics of workloads interact with each other, we conducted a Spearman's correlation analysis and identified characteristics that exhibit a correlation greater than 0.25. The stateless nature of workloads shows a significant correlation with other workload characteristics (Figure 6). Most stateless workloads are easier to preempt as they can recover from faults or failures. In addition, they can migrate to other regions without causing any data loss or corruption as they do not rely on any local data or state that is tied to their region of deployment.

The results on the correlation between workload characteristics from our empirical study can further scale out to cover a greater number of workloads, workload characteristics and runtime signals with the help of applying machine learning models. This will be part of our future work.

## 6 Future Work

In this work, we conducted a large-scale empirical study of real-world workloads to propose a framework for a comprehensive characterization of cloud workloads. We also showed the interplay between workload characteristics and the appli-

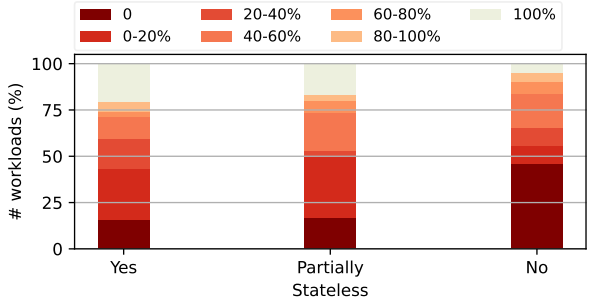

(a) Stateless nature and preemptibility of workloads.

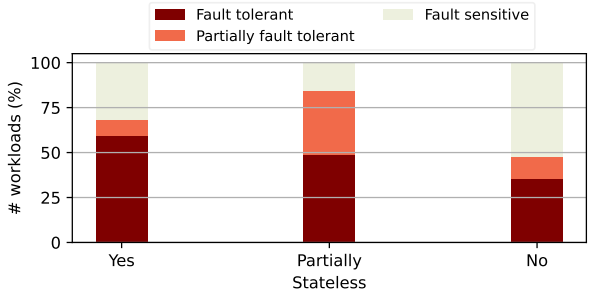

(b) Stateless nature and fault tolerance of workloads.

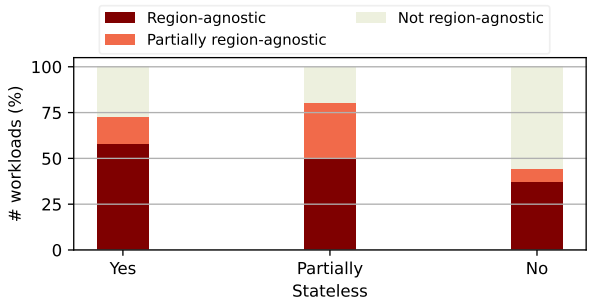

(c) Stateless nature and region-agnostic nature of workloads.

Figure 6: Correlation between workload characteristics.

cability of various cloud optimizations. However, we believe that this is just a first step in the direction of building *self-optimizing* clouds, which can leverage workload knowledge to maximize efficiency while guaranteeing reliability. We see several opportunities for future research in terms of automating the workload characterization and, also, evolving the existing cloud infrastructure to be workload aware.

**Workload characterization and modeling.** In this work, we used a survey to collect information about the workloads from the workload owners and other metadata sources. However, this is not a scalable approach as we were able to gather data for workloads out of the thousands of workloads at Microsoft. Further, a manual approach would not work for third party workloads where the cloud providers might not have direct access to the workload owners and metadata. Hence, there is a need to automate the workload characterization process to be able to scale it to a greater number of workloads and, also, to third party workloads. As a next step, we plan to use state-of-the-art approaches in the machine learning domain for sample-efficient learning along with the workload characterization from this study. For example, few-shot learning techniques [53] have been shown to leverage a small amount of labelled data to efficiently model different classes by learning similarity (and dis-similarity).

Further, the characterization of workloads can be enhanced by incorporating additional signals, such as deployment information, service level objectives, resource utilization and allocation metrics, as well as upstream dependencies. To achieve this, it is necessary to construct a comprehensive representation of the feature space utilizing all available signals.

**Enabling cloud optimizations in production.** Today, cloud providers like AWS and Azure are operating hyper-scale clouds and must manage the trade-off between maximizing efficiency (hence, profits) and reliability. For this purpose, cloud providers introduce new VM types (such as Spot VMs) and optimizations (such as auto-scaling, overclocking) to maximize utilization of the cloud resources. These optimizations require knowledge of workload characteristics, which are not readily available, hence prompting this study. At Microsoft, we are starting to leverage the insights from this study to optimize the capacity allocation and utilization. For instance, we have leveraged the understanding of geographical constraints of workloads to migrate region-agnostic workloads from cloud regions with capacity constraints to other regions. Eventually, we want to scale it to other optimizations, and also do this for third-party workloads.

**Need for a cloud-workload interface.** To truly accelerate the adoption of cloud optimizations, we need a hybrid approach where we not only automate the workload characterization but also enable workload owners to provide knowledge and hints about their workloads, which cloud providers can use to enable the optimizations. To this end, we envision building an extensible interface between the workload and cloud providers, which will allow them to communicate workload characteristics, some of which can dynamically change over time. Through conveying these workload details, cloud optimizations can be enabled to maximize efficiency while ensuring reliability.

## 7 Related Work

**Resource management and ML-centric approaches.** Some works optimize resource utilization while meeting performance using limited workload information such as QoS constraints or resource usage [2, 22, 23, 30, 68]. Other works [33, 40, 65] propose optimizations that assume knowledge of the workload characteristics. There are also ML-centric approaches exploring how and where ML should be infused in cloud platforms [12, 20, 59]. Our work is complimentary to these efforts as our insights enable these approaches and motivate future research into improving existing resource management/optimizations.

**Workload profiling and characterization.** Related works in this area often study workload characteristics related to resource utilization and workload deployment such as lifetime or task duration [14–16, 20, 24, 26, 31, 36–39, 43, 48, 50, 67] and explore the heterogeneity in those characteristics [35, 37, 44, 47]. These works provide insights on the heterogeneity and disparity in the characteristics of workload deployment and resource utilization.

Other studies explore workload characteristics including failure distribution, correlation, arrival rate, interference between resources [28, 56], volatility of resource demand, usage by region and customer segments [55], and microservice dependencies [32, 38]. Unlike other large-scale studies, such as Borg [56] that target workloads deployed on specific clusters, this work examines the characteristics of workloads that are deployed across 49 regions worldwide and used by millions of users.

The closest literature to our work are [45, 62], which adopt a data-driven approach to identify the opportunities and design challenges to enable ML inference locally on smartphones and other edge platforms. The work studies ML workloads based on the signals available (*e.g.*, inference time) and discusses the implications of the studied characteristics on design decisions.

These works often study workload signals or characteristics directly derived from a single workload signal/telemetry. Unfortunately, the workload characteristics relevant to cloud platform optimizations are often complex and depend on multiple workload signals. These characteristics are not directly evident from the workload signals available from the cloud platform. Our work focuses on uncovering such fundamental workload characteristics and deriving insights for reliability and efficiency improvements.

## 8 Conclusion

In this work, we conduct the first large-scale study of first-party workloads at Microsoft. This study provides a comprehensive understanding of the characteristics of first-party workloads and their variations across multiple workloads. The study also identifies unique workload characteristics and suitable workloads for each cloud platform optimization. The findings of this study can be leveraged to improve the efficiency and reliability of cloud platforms. These findings can also assist cloud users in decreasing the overall cost and enhancing performance. Additionally, we highlight future research opportunities in this field. Overall, this study contributes to the ongoing effort to optimize cloud services by providing a deeper understanding of workload characteristics and their impact on cloud performance.

## References

[1] Muhammad Abdullah Adnan, Ryo Sugihara, and Rajesh K Gupta. Energy efficient geographical load balancing via dynamic deferral of workload. In *CLOUD*, 2012.

[2] Omid Alipourfard, Hongqiang Harry Liu, Jianshu Chen, Shivaram Venkataraman, Minlan Yu, and Ming Zhang. CherryPick: Adaptively Unearthing the Best Cloud Configurations for Big Data Analytics. In *NSDI*, 2017.

[3] Amazon. Amazon Web Services, 2023. https://aws.amazon.com/.

[4] Amazon Elastic Compute Cloud. Amazon EC2 Spot Instances, 2019. https://aws.amazon.com/ec2/spot/.

[5] Amazon Elastic Compute Cloud. Running batch jobs at scale for less, 2020. https://aws.amazon.com/getting-started/hands-on/run-batch-jobs-at-scale-with-ec2-spot/.

[6] Amazon Web Services. Amazon EC2 Auto Scaling, 2023. https://aws.amazon.com/ec2/autoscaling/.

[7] Pradeep Ambati, Íñigo Goiri, Felipe Frujeri, Alper Gun, Ke Wang, Brian Dolan, Brian Corell, Sekhar Pasupuleti, Thomas Moscibroda, Sameh Elnikety, Marcus Fontoura, and Ricardo Bianchini. Providing SLOs for Resource-Harvesting VMs in Cloud Platforms. In *OSDI*, 2020.

[8] Microsoft Azure. Introducing B-Series, Our New Burstable VM Size, 2019. https://azure.microsoft.com/en-us/blog/introducing-b-series-our-new-burstable-vm-size/.

[9] Microsoft Azure. Use low-priority VMs with Batch, 2019. https://docs.microsoft.com/en-us/azure/batch/batch-low-pri-vms.

[10] Microsoft Azure. Azure Spot Virtual Machines, 2020. https://azure.microsoft.com/en-us/pricing/spot.

[11] Ataollah Fatahi Baarzi and George Kesidis. Showar: Right-sizing and efficient scheduling of microservices. In *SoCC*, 2021.

[12] Ricardo Bianchini, Marcus Fontoura, Eli Cortez, Anand Bonde, Alexandre Muzio, Ana-Maria Constantin, Thomas Moscibroda, Gabriel Magalhaes, Girish Bablani, and Mark Russinovich. Toward ML-Centric Cloud Platforms. *Communications of the ACM*, 2020.

[13] BMC Blogs. Service availability, 2020. https://www.bmc.com/blogs/service-availability-calculation-metrics/.

[14] Wenyan Chen, Kejiang Ye, Yang Wang, Guoyao Xu, and Cheng-Zhong Xu. How does the workload look like in production cloud? Analysis and clustering of workloads on alibaba cluster trace. In *ICPADS*, 2018.

[15] Yanpei Chen, Archana Sulochana Ganapathi, Rean Griffith, and Randy H Katz. Analysis and Lessons from a Publicly Available Google Cluster Trace. *EECS Department, University of California, Berkeley, Tech. Rep. UCB/EECS-2010-95*, 94, 2010.

[16] Yue Cheng, Zheng Chai, and Ali Anwar. Characterizing co-located datacenter workloads: An alibaba case study. In *Asia-Pacific Workshop on Systems*, 2018.

[17] Amazon Elastic Compute Cloud. Burstable Performance Instances, 2019. https://docs.aws.amazon.com/AWSEC2/latest/UserGuide/burstable-performance-instances.html.

[18] Google Cloud. Preemptible VM Instances, 2020. https://cloud.google.com/compute/docs/instances/preemptible.

[19] Jacob Cohen. A coefficient of agreement for nominal scales. *Educational and psychological measurement*, 20(1):37–46, 1960.

[20] Eli Cortez, Anand Bonde, Alexandre Muzio, Mark Russinovich, Marcus Fontoura, and Ricardo Bianchini. Resource Central: Understanding and predicting workloads for improved resource management in large cloud platforms. In *SOSP*, 2017.

[21] Konrad Cłapa and Brian Gerrard. *FProfessional Cloud Architect Google Cloud Certification Guide*. Packt Publishing, 2021.

[22] Christina Delimitrou and Christos Kozyrakis. Paragon: QoS-aware scheduling for heterogeneous datacenters. In *ASPLOS*, 2013.

[23] Christina Delimitrou and Christos Kozyrakis. Quasar: Resource-efficient and qos-aware cluster management. In *ASPLOS*, 2014.

[24] Sheng Di, Derrick Kondo, and Franck Cappello. Characterizing cloud applications on a Google data center. In *ICPP*, 2013.

[25] Alexander Fuerst, Stanko Novaković, Íñigo Goiri, Gohar Irfan Chaudhry, Prateek Sharma, Kapil Arya, Kevin Broas, Eugene Bak, Mehmet Iyigun, and Ricardo Bianchini. Memory-harvesting VMs in Cloud Platforms. In *ASPLOS*, 2022.

[26] Peter Garraghan, Paul Townend, and Jie Xu. An Analysis of the Server Characteristics and Resource Utilization in Google Cloud. In *IC2E*. IEEE, 2013.

[27] Google. Google Cloud, 2023. https://cloud.google.com/.

[28] Jing Guo, Zihao Chang, Sa Wang, Haiyang Ding, Yihui Feng, Liang Mao, and Yungang Bao. Who limits the resource efficiency of my datacenter: An analysis of Alibaba datacenter traces. In *IWQoS*, 2019.

[29] Majid Jalili, Ioannis Manousakis, Íñigo Goiri, Pulkit A Misra, Ashish Raniwala, Husam Alissa, Bharath Ramakrishnan, Phillip Tuma, Christian Belady, Marcus Fontoura, et al. Cost-efficient overclocking in immersion-cooled datacenters. In *ISCA*, 2021.

[30] Seyyed Ahmad Javadi, Amoghavarsha Suresh, Muhammad Wajahat, and Anshul Gandhi. Scavenger: A black-box batch workload resource manager for improving utilization in cloud environments. In *SoCC*, 2019.

[31] Congfeng Jiang, Guangjie Han, Jiangbin Lin, Gangyong Jia, Weisong Shi, and Jian Wan. Characteristics of co-allocated online services and batch jobs in internet data centers: a case study from Alibaba cloud. *IEEE Access*, 7:22495–22508, 2019.

[32] Cinar Kilcioglu, Justin M Rao, Aadharsh Kannan, and R Preston McAfee. Usage patterns and the economics of the public cloud. In *WWW*, 2017.

[33] Alok Gautam Kumbhare, Reza Azimi, Ioannis Manousakis, Anand Bonde, Felipe Frujeri, Nithish Mahalingam, Pulkit A Misra, Seyyed Ahmad Javadi, Bianca Schroeder, Marcus Fontoura, et al. Prediction-Based Power Oversubscription in Cloud Platforms. In *USENIX ATC*, 2021.

[34] Nuno Laranjeiro and Marco Vieira. Towards fault tolerance in web services compositions. In *Proceedings of the 2007 Workshop on Engineering Fault Tolerant Systems*, page 2–8, 2007.

[35] Bingwei Liu, Yinan Lin, and Yu Chen. Quantitative workload analysis and prediction using google cluster traces. In *INFOCOM Workshops*, 2016.

[36] Zitao Liu and Sangyeun Cho. Characterizing machines and workloads on a Google cluster. In *ICPP Workshops*, 2012.

[37] Chengzhi Lu, Kejiang Ye, Guoyao Xu, Cheng-Zhong Xu, and Tongxin Bai. Imbalance in the cloud: An analysis on alibaba cluster trace. In *Big Data*, 2017.

[38] Shutian Luo, Huanle Xu, Chengzhi Lu, Kejiang Ye, Guoyao Xu, Liping Zhang, Yu Ding, Jian He, and Chengzhong Xu. Characterizing microservice dependency and performance: Alibaba trace analysis. In *SoCC*, 2021.

[39] Ashraf Mahgoub, Edgardo Barsallo Yi, Karthick Shankar, Eshaan Minocha, Sameh Elnikety, Saurabh Bagchi, and Somali Chaterji. WISEFUSE: Workload Characterization and DAG Transformation for Serverless Workflows. *POMACS*, 2022.

[40] Ming Mao and Marty Humphrey. Auto-scaling to minimize cost and meet application deadlines in cloud workflows. In *SC*, 2011.

[41] Microsoft. Azure, 2023. https://azure.microsoft.com/.

[42] Microsoft Azure. Overview of autoscale in Microsoft Azure, 2022. https://docs.microsoft.com/en-us/azure/azure-monitor/autoscale/autoscale-overview.

[43] Asit K Mishra, Joseph L Hellerstein, Walfredo Cirne, and Chita R Das. Towards characterizing cloud backend workloads: insights from Google compute clusters. *SIGMETRICS*, 37(4):34–41, 2010.

[44] Md Rasheduzzaman, Md Amirul Islam, Tasvirul Islam, Tahmid Hossain, and Rashedur M Rahman. Task shape classification and workload characterization of google cluster trace. In *IACC*, 2014.

[45] Vijay Janapa Reddi, Christine Cheng, David Kanter, Peter Mattson, Guenther Schmuelling, Carole-Jean Wu, Brian Anderson, Maximilien Breughe, Mark Charlebois, William Chou, et al. MLPerf inference benchmark. In *ISCA*, 2020.

[46] Benjamin Reidys, Jinghan Sun, Anirudh Badam, Shadi Noghabi, and Jian Huang. BlockFlex: Enabling Storage Harvesting with Software-Defined Flash in Modern Cloud Platforms. In *OSDI*, 2022.

[47] Charles Reiss, Alexey Tumanov, Gregory R Ganger, Randy H Katz, and Michael A Kozuch. Heterogeneity and dynamicity of clouds at scale: Google trace analysis. In *SoCC*, 2012.

[48] Charles Reiss, Alexey Tumanov, Gregory R Ganger, Randy H Katz, and Michael A Kozuch. Towards understanding heterogeneous clouds at scale: Google trace analysis. *Intel Science and Technology Center for Cloud Computing, Tech. Rep*, 84:1–21, 2012.

[49] Amazon Web Services. Processor state control for your EC2 instance, 2023. https://docs.aws.amazon.com/AWSEC2/latest/UserGuide/processor_state_control.html.

[50] Mohammad Shahrad, Rodrigo Fonseca, Íñigo Goiri, Gohar Chaudhry, Paul Batum, Jason Cooke, Eduardo Laureano, Colby Tresness, Mark Russinovich, and Ricardo Bianchini. Serverless in the wild: Characterizing and optimizing the serverless workload at a large cloud provider. In *USENIX ATC*, 2020.

[51] Mohammad Shahrad and David Wentzlaff. Availability knob: Flexible user-defined availability in the cloud. In *SoCC*, 2016.

[52] Jiuchen Shi, Kaihua Fu, Quan Chen, Changpeng Yang, Pengfei Huang, Mosong Zhou, Jieru Zhao, Chen Chen, and Minyi Guo. Characterizing and orchestrating VM reservation in geo-distributed clouds to improve the resource efficiency. In *SoCC*, 2022.

[53] Jake Snell, Kevin Swersky, and Richard Zemel. Prototypical networks for few-shot learning. *Advances in neural information processing systems*, 30, 2017.

[54] Anselm Strauss and Juliet M Corbin. *Grounded theory in practice*. Sage, 1997.

[55] Huangshi Tian, Yunchuan Zheng, and Wei Wang. Characterizing and synthesizing task dependencies of data-parallel jobs in alibaba cloud. In *SoCC*, 2019.

[56] Muhammad Tirmazi, Adam Barker, Nan Deng, Md E Haque, Zhijing Gene Qin, Steven Hand, Mor Harchol-Balter, and John Wilkes. Borg: The Next Generation. In *EuroSys*, 2020.

[57] Astrid Undheim, Ameen Chilwan, and Heegaard. Differentiated availability in cloud computing slas. In *Proceedings of the 2011 IEEE/ACM 12th International Conference on Grid Computing*, page 129–136, 2021.

[58] Cheng Wang, Bhuvan Urgaonkar, Neda Nasiriani, and George Kesidis. Using burstable instances in the public cloud: Why, when and how? 2017.

[59] Yawen Wang, Kapil Arya, Marios Kogias, Manohar Vanga, Aditya Bhandari, Neeraja J Yadwadkar, Siddhartha Sen, Sameh Elnikety, Christos Kozyrakis, and Ricardo Bianchini. SmartHarvest: Harvesting idle CPUs safely and efficiently in the cloud. In *EuroSys*, 2021.

[60] Philipp Wiesner, Ilja Behnke, Dominik Scheinert, Kordian Gontarska, and Lauritz Thamsen. Let's wait awhile: how temporal workload shifting can reduce carbon emissions in the cloud. In *Middleware*, 2021.

[61] Dan Williams, Hani Jamjoom, Yew-Huey Liu, and Hakim Weatherspoon. Overdriver: Handling Memory Overload in an Oversubscribed Cloud. *VEE*, 2011.

[62] Carole-Jean Wu, David Brooks, Kevin Chen, Douglas Chen, Sy Choudhury, Marat Dukhan, Kim Hazelwood, Eldad Isaac, Yangqing Jia, Bill Jia, et al. Machine learning at Facebook: Understanding inference at the edge. In *HPCA*, 2019.

[63] Randolph Yao, Chuan Luo, Bo Qiao, Qingwei Lin, Tri Tran, Gil Lapid Shafriri, Yingnong Dang, Raphael Ghelman, Pulak Goyal, Eli Cortez, et al. Infusing ML into VM Provisioning in Cloud. In *CloudIntelligence*, 2021.

[64] Randolph Yao, Chuan Luo, Bo Qiao, Qingwei Lin, Tri Tran, Gil Lapid Shafriri, Yingnong Dang, Raphael Ghelman, Pulak Goyal, Eli Cortez, Daud Howlader, Sushant Rewaskar, Murali Chintalapati, and Dongmei Zhang. Infusing ML into VM Provisioning in Cloud. In *CloudIntelligence*, 2021.

[65] Chaojie Zhang, Alok Gautam Kumbhare, Ioannis Manousakis, Deli Zhang, Pulkit A Misra, Rod Assis, Kyle Woolcock, Nithish Mahalingam, Brijesh Warrier, David Gauthier, et al. Flex: High-availability datacenters with zero reserved power. In *ISCA*, 2021.

[66] Chengliang Zhang, Minchen Yu, Wei Wang, and Feng Yan. MArk: Exploiting Cloud Services for Cost-Effective, SLO-Aware Machine Learning Inference Serving. In *USENIX ATC*, 2019.

[67] Qi Zhang, Joseph Hellerstein, and Raouf Boutaba. Characterizing Task Usage Shapes in Google Compute Clusters. 2011.

[68] Yanqi Zhang, Weizhe Hua, Zhuangzhuang Zhou, G Edward Suh, and Christina Delimitrou. Sinan: ML-based and QoS-aware resource management for cloud microservices. In *ASPLOS*, 2021.

## A Survey Questions

1. What is the name and service Id of your service?

2. What are the main functions of your service? Please share documents/wiki/ppt describing the main functions of your service, if any. [2-3 sentences]

3. Is your service composed of multiple services or components? If yes, please list the name / brief description of these services or components.

4. We define a service as customer facing if it handles real time customer traffic. Is your service customer facing?

    (a) Yes    (b) No

5. Patterns in CPU utilization can differ between customer facing and non-customer facing services. Are there any other signals or data sources for your service that would reveal whether it is customer facing or not? [1-2 sentences]

6. Please specify the approximate expected availability of your service.

| Five nines | Four nines | Three nines | Two nines | One nine | None |
|---|---|---|---|---|---|
| (a) | (b) | (c) | (d) | (e) | (f) |

7. Do you have strict requirements on time it takes to deploy your service?

    (a) Yes    (b) No

8. If yes, what is the maximum allowed time to deploy VM (virtual machines)? Please specific the approximate value (in sec or min or hour or days)

9. Delay tolerance of a service defines the target/expected performance for a service. A service is delay tolerant if it can tolerate delay (relatively insensitive to higher latency). Is your service delay tolerant?

    (a) Yes    (b) No

10. If yes, please specify approximate latency SLA. Where can we get the telemetry of your service's latency?

11. A stateless service requires no past data nor state to be stored or persisted. For instance, a web server processing an independent request without retrieving any kind of application context or state is a stateless service. Is your service stateless?

    (a) Yes    (b) No    (c) Partially

12. A preemptible service can withstand loss of its VM instances and can be resumed later. Given that services in general do not support losing more than a certain percentage of its VMs, what percentage of your service is preemptible?

| 0 | 0-20% | 20-40% | 40-60% | 60-80% | 80-100% | 100% |
|---|---|---|---|---|---|---|
| (a) | (b) | (c) | (d) | (e) | (f) | (g) |

13. If your service is preemptible, what is the maximum amount of interrupt you can withstand before negative customer impact? Please specify the approximate value (in sec or min or hour)?

14. A service is fault-tolerant if its performance is insensitive in reaction to system changes, e.g., a failing computation node is not impacting on latency. Is your service fault-tolerant?

    (a) Yes    (b) No    (c) Partially

15. Region-Agnostic Workloads (RAW) are those that could be deployed or migrated at least to any other region within a certain geo-locale without any negative impact to its functioning. Is your service region-agnostic?

    (a) Yes  (b) No  (c) Partially

16. Please specify the approximate percentage of your service which is not RAW.

17. If your service or part of it is not RAW, what sticky design patterns (multiple choices, such as IP ranges, Storage affinities, Local caches, etc.) prevent them from shifting to other regions?

    (a) IP ranges  (b) Storage affinities  (c) Local cache  (d) Other, please specify

18. Please specify any data sources or documents from which we can get to know more about some of the above specified sticky design patterns. [1-3 sentences]

19. If we want to understand more about the basic properties of your service, telemetry, incident reports, service level objectives, etc. what are the relevant data sources to look at?

20. Are there any other characteristics of your service which would help improve its efficiency and reliability?

21. If you have any questions or difficulties in answering any of the questions given in this survey, please specify here.